\documentclass[reprint, aps, prd, superscriptaddress]{revtex4-2}

\usepackage{amsmath}
\usepackage{amssymb}
\usepackage{amsfonts}
\usepackage{graphicx}
\usepackage[utf8]{inputenc}
\usepackage[switch, columnwise]{lineno}

\usepackage{multirow}
\usepackage{booktabs}		
\usepackage[dvipsnames]{xcolor}
\usepackage{natbib}
\usepackage{array}
\usepackage{float}

\newcolumntype{a}[1]{>{\raggedleft\arraybackslash}p{#1}}

\usepackage[margin=2cm]{geometry}   
\graphicspath{{figures/}}			
\numberwithin{equation}{section}    
\usepackage[nolist]{acronym}

\usepackage{hyperref}

\newcommand{\diff}{\mathop{}\!\mathrm{d}}

\begin{document}

\title{Efficient sampling of constrained high-dimensional theoretical spaces\\ with machine learning}
\date{\today}
\author{Jacob Hollingsworth}
\affiliation{Department of Physics and Astronomy, University of California, Irvine, CA 92697}
\author{Michael Ratz}
\affiliation{Department of Physics and Astronomy, University of California, Irvine, CA 92697}
\author{Philip Tanedo}
\affiliation{Department of Physics and Astronomy, University of California, Riverside, California 92521}
\author{Daniel Whiteson}
\affiliation{Department of Physics and Astronomy, University of California, Irvine, CA 92697}

\begin{abstract}
\noindent Models of physics beyond the Standard Model often contain a large number of parameters. These form a high-dimensional space that is computationally intractable to fully explore.  Experimental constraints project onto a subspace of viable parameters, but mapping these constraints to the underlying parameters is also typically intractable. Instead, physicists often resort to scanning small subsets of the full parameter space and testing for experimental consistency. We propose an alternative approach that uses generative models to significantly improve the computational efficiency of sampling high-dimensional parameter spaces. To demonstrate this, we sample the constrained and phenomenological Minimal Supersymmetric Standard Models subject to the requirement that the sampled points are consistent with the measured Higgs boson mass. Our method achieves orders of magnitude improvements in sampling efficiency compared to a brute force search. 
\end{abstract}

\maketitle

\section{Introduction}

Models of physics beyond the Standard Model often feature many new parameters that are unknown \emph{a priori} and may only be determined by experiment. However, experimental constraints are not trivial to apply, as they often are expressed in terms of weak-scale observables rather than the theory's fundamental parameters.  While it is often straightforward (if computationally expensive) to calculate the weak-scale observables from the parameters, the inverse problem is typically intractable. That is, weak scale constraints do not allow for a trivial reduction of the dimensionality of the theory space.

 The standard approach is to numerically scan over the theoretical parameters and reject those that are not consistent with experimental data. However, the number of samples required for a brute-force search of the parameter space increases exponentially with its dimension. Thus, particle physicists studying models of new physics are often faced with a computationally intractable task. Out of pragmatism, one then restricts to a more tractable subset of parameters based on theoretical prejudice. The danger of this approach is that one may miss viable parameters that are both consistent with experimental observations and generate novel phenomenology.

The \ac{MSSM} is a well-known example new physics model with a large number of free parameters ($\sim 100$); most of which are the masses and couplings of the supersymmetric partners of Standard Model particles~\cite{Martin1997}. This overwhelming dimensionality prohibits a fully general survey of the parameter space. Studies of the \ac{MSSM} typically restrict to theoretically motivated subspaces~\cite{Cohen_2013, Ghosh_2012, Han_2017, Allanach_2011, Buchmueller2014, Bridges_2011, PhysRevD.91.055002, cahillrowley2013pmssm, Aad_2015, Khachatryan_2016, Caron2017, Kronheim_2021}. These include the 4+1 dimensional \ac{cMSSM} as well as the 19 dimensional \ac{pMSSM} \cite{PhysRevLett.49.970, Djouadi1998d}. However, even these reduced spaces are difficult to scan using a brute-force search.

High dimensionality is not the only challenge when scanning the parameters of the \ac{MSSM}. The fundamental parameters of the theory are defined at some high energy scale---e.g.~the scale of a so-called \ac{GUT}---and must be evolved to the energy scale of the experiment. This evolution requires one to solve the coupled \acp{RGE} for the high-scale parameters over many orders of magnitude to the weak scale.
The computational cost of \ac{RGE} running and calculating experimental observables for a single set of parameters is expensive, $\mathcal{O}(\text{second})$ for a modern CPU. 

Many recent scans have incorporated machine learning in some capacity to decrease the computational burden of brute-force searching these spaces \cite{Caron2017, Kronheim_2021, Bridges_2011}. These use various machine learning models to learn the \emph{forward} problem of determining weak scale properties given \ac{GUT} scale parameters. 
This bypasses the need to perform \ac{RGE} running and weak scale computations, however one is still faced with the challenge of doing a brute-force search over a high-dimensional parameter space. Machine learning models for the forward problem are thus only a constant improvement compared to the exponential dependence on the dimension of the space.

In this work, we introduce two methods to efficiently sample high-dimensional parameter spaces subject to constraints at the weak scale. We test these frameworks by sampling regions of the \ac{cMSSM} and \ac{pMSSM} parameter spaces that admit a Higgs mass consistent with its experimental value \cite{Aad_2012, Chatrchyan_2012}. The first uses a deep neural network to machine-learn the likelihood of an event satisfying this constraint and then samples this likelihood using \ac{HMC}. The second trains a generative model known as a normalizing flow. We then compare the performance of these frameworks to random sampling.

These methods allows us to directly and quickly generate points in the parameter space that admit a consistent Higgs mass. By solving the inverse problem of sampling \ac{GUT} scale parameters given weak scale properties, we aim to minimize inefficiencies that arise in a brute-force search. 

Our presentation is a proof of concept for these generative models and is encouraging for practical applications. For example, the ability to efficiently scan the \ac{MSSM} parameter space makes it much easier to determine the high-scale parameters that are consistent with a new particle's mass and width if a sparticle is discovered. Alternatively, a trained generative model may permit scans over parameters that are consistent with experimental observations to search for specific theoretical features that one may wish to study, for example: gauge coupling unification, a particular type of dark matter particle, or low fine tuning measures.

As a demonstration of the efficiency of the generative models, we scan the \ac{cMSSM} and \ac{pMSSM} parameter spaces for points that produce the Higgs mass and that saturate the observed dark matter relic density, requiring \cite{Spergel_2003, Bennett_2013}
\begin{linenomath*}
\[
\begin{array}{r@{{}}c@{{}}l}
122~\text{GeV}  <{} &m_h &{}< 128~\text{GeV}\; ,\\
0.08 < {}&\Omega_{\textrm{DM}}h^2 &{}< 0.14\;.
\end{array}\]
\end{linenomath*}
In this study, the generative models have been trained for consistency with the Higgs mass, not the relic density. We compare a brute-force scan using random sampling to a generative model that has been trained to sample points that admit a consistent Higgs mass. We show that the generative models dramatically increase the sampling efficiency of this scan. 

\section{Methods}
\subsection{Data Generation}

The \ac{cMSSM} contains 4 continuous parameters defined at the \ac{GUT} scale and 1 discrete sign parameter.
These are the universal scalar mass $m_0$, the universal gaugino mass $M_{1/2}$, universal trilinear coupling $A_0$, the ratio of Higgs vacuum expectation values $\tan\beta$, and the sign of $\mu$. 
The \ac{pMSSM} is the most general subspace of the \ac{MSSM} that admits first and second generation universality, no new sources of CP violation, and no flavor changing neutral currents \cite{Djouadi1998d}. The full list parameters of the \ac{pMSSM} are listed as part of Table \ref{pmssm_params}. Upon evolving these parameters to the weak scale, one is able to calculate observable quantities such as the Higgs mass and the dark matter thermal relic density, among others.

Our datasets are formed by uniform random sampling within bounded regions of the parameter space at the \ac{GUT} scale. These bounds are listed for the \ac{cMSSM} and the \ac{pMSSM} in Table \ref{cmssm_params} and Table \ref{pmssm_params}, respectively \cite{Cohen_2013, cahillrowley2013pmssm}, and are chosen to cover large volumes of the parameter space that are sensitive to modern collider experiments. For the \ac{cMSSM}, we fix $\operatorname{sign}(\mu)=1$. We sample approximately $1.5\times 10^6$ datapoints in the \ac{cMSSM} and approximately $1.95 \times 10^7$ datapoints in the pMSSM. Once sampled, we calculate Higgs masses and relic densities with micrOMEGAs, which internally uses SoftSUSYv4.1.0 to perform \ac{RGE} running \cite{belanger2010micromegas, Allanach_2002}.

\begin{table}
\centering
\begin{tabular}{ l  l  l}\toprule
Parameter & Domain  & Description \\ \midrule
$m_0$ & $[0, 10]$ TeV & Universal scalar mass \\
$m_{1/2}$ & $[0,10]$ TeV & Universal gaugino mass\\
$A_0$ & $[-6 m_0, 6 m_0]$ TeV & Universal trilinear coupling\\
$\tan\beta$ & $[1.5,50]$ & Ratio of Higgs VEVs\\ 
\bottomrule
\end{tabular}
\caption{Parameter bounds in the \ac{cMSSM} scan, following Ref.~\cite{Cohen_2013}. A uniform prior is used for all parameters except $A_0$,  where we uniformly sample $A_0 / m_0$. 
}\label{cmssm_params}
\end{table}

\begin{table}
\centering
\begin{tabular}{ l  l  l} \toprule
Parameter & Domain & Description \\ \midrule
$|M_{1}|$ & $[.05, 4]$ TeV& Bino mass \\
$|M_{2}|$ & $[.1,4]$ TeV& Wino mass\\
$M_{3}$ & $[.4, 4]$ TeV& Gluino mass \\
$|\mu|$ & $[.1,4]$ TeV& Bilinear Higgs mass\\
$|A_t|$  & $[0, 4]$ TeV& Trilinear top coupling \\
$|A_b|$ & $[0,4]$ TeV& Trilinear bottom coupling\\
$|A_{\tau}|$ & $[0, 4]$ TeV& Trilinear $\tau$ coupling \\
$M_A$ & $[.1, 4]$ TeV& Pseudo-scalar Higgs mass \\
$m_{\tilde{L}_1}$ & $[.1, 4]$ TeV& 1st gen.~l.h.~slepton mass \\
$m_{\tilde{e}_1}$ & $[.1, 4]$ TeV& 1st gen.~r.h.~slepton mass \\
$m_{\tilde{L}_3}$ & $[.1, 4]$ TeV& 3rd gen.~l.h.~slepton mass \\
$m_{\tilde{e}_3}$ & $[.1, 4]$ TeV& 3rd gen.~r.h.~slepton mass\\
$m_{\tilde{Q}_1}$ & $[.4, 4]$ TeV& 1st gen.~l.h.~squark mass \\
$m_{\tilde{u}_1}$ & $[.4, 4]$ TeV& 1st gen.~r.h.~$u$-type squark mass\\
$m_{\tilde{d}_1}$ & $[.4, 4]$ TeV& 1st gen.~r.h.~$d$-type squark mass \\
$m_{\tilde{Q}_3}$ & $[.2, 4]$ TeV& 3rd gen.~l.h.~squark mass \\
$m_{\tilde{u}_3}$ & $[.2, 4]$ TeV& 3rd gen.~r.h.~$u$-type squark mass\\
$m_{\tilde{d}_3}$ & $[.2, 4]$ TeV & 3rd gen.~r.h.~$d$-type squark mass \\
$\tan\beta$ & $[1,60]$ & Ratio of Higgs VEVs\\ \bottomrule
\end{tabular}
\caption{Parameter bounds in the \ac{pMSSM} scan, following  Ref.~\cite{cahillrowley2013pmssm}. A uniform prior is used for all parameters. 
``Left-handed'' and ``right-handed'' are abbreviated by l.h.\ and r.h., respectively.}\label{pmssm_params}
\end{table}

We apply the following two theoretical constraints:
the parameters allow for consistent electroweak symmetry breaking and that all squared masses are positive. In addition to these, we also require that SoftSUSY converges. We do not require that the lightest supersymmetric particle is neutral, though this is the case for 90\% of the cMSSM and 99\% of the pMSSM parameter points with a consistent Higgs mass. 

The theoretical uncertainty in Higgs mass calculations is significantly larger than its experimental uncertainty \cite{Athron_2017}. We take the uncertainty in the Higgs mass calculations to be $\sigma_{m_h} = 3~$GeV for all points in the data set \cite{Cohen_2013, cahillrowley2013pmssm}.

\subsection{Neural Network}
To train the neural network, we assign all points in the dataset a likelihood
\begin{linenomath*}
\begin{align}
L(\theta) = \begin{cases}
1 & |m_h(\theta) - m_{h,\mathrm{exp}}| < \sigma_{m_h}\;, \\ 
0 & \text{otherwise}\;,
\end{cases}
\end{align}
\end{linenomath*}
where the normalization constant is ignored. All data points that fail the theoretical constraints are assigned a likelihood of zero.

We use a deep neural network to learn the function $L(\theta)$ \cite{mlapr}.
This has two benefits. First, it greatly reduces the time required to evaluate the likelihood of a point. Second, it provides a differentiable interpolation of $L(\theta)$. In the next section we show that \ac{HMC} requires a significant number of likelihood evaluations as well as gradients of the likelihood. It thus utilizes the full potential of these benefits. 

We train a deep neural network $\hat{L}(\theta)$ to minimize the usual L2 loss function
\begin{linenomath*}
\begin{align}
\mathcal{L} = |\hat{L}(\theta) - L(\theta)|^2\;.
\end{align}
\end{linenomath*}
We use a training, validation, and testing split of $0.7, 0.15, 0.15$ respectively for both datasets. Batch norm and dropout layers are used in between each hidden layer of the neural network. Backpropogation is performed using the ADAM optimizer \cite{kingma2017adam}.

Some of the \ac{pMSSM} parameters in Table~\ref{pmssm_params} span a disconnected range of positive and negative values, for example $M_1$, $M_2$ and $\mu$. We preprocess these parameters by shifting negative values to create a single continuous domain; for example, for $\mu$ we shift the negative values by 200 GeV. This has no physical significance and simply prepares the data for input into the neural network.
We then standardize each feature. For the \ac{cMSSM} dataset, we use the feature $A_0 / m_0$ in place of $A_0$, as this feature is uniformly distributed.

\subsection{Hamiltonian Monte Carlo}

The Hamiltonian Monte Carlo method is a Markov chain Monte Carlo technique that allows distant proposals with high acceptance rates \cite{neal2012mcmc, betancourt2018conceptual}. First, we define an auxiliary momentum variable $p$, where each component is initially drawn from a normal distribution. Next, we define a potential energy function given by
\begin{linenomath*}
\begin{align}
V(\theta) = -\log(\hat{L}(\theta))\;.
\end{align}  
\end{linenomath*}
The kinetic energy function takes the familiar form with unit mass $m=1$,
\begin{linenomath*}
\begin{align}
T=\frac{1}{2} p^2\;.
\end{align}
\end{linenomath*}
We then evolve the system from time $t=0$ to $t=\tau$ by solving Hamilton's equations of motion
\begin{linenomath*}
\begin{align}
\frac{\diff\theta_i}{\diff t} &= p_i\;,
&
\frac{\diff p_i}{\diff t} &= \frac{\nabla \hat{L}(\theta)}{\hat{L}(\theta)}
\; .
\end{align}
\end{linenomath*}
We solve the equations of motion using the leap-frog algorithm so that energy is approximately conserved. We take $\theta(\tau)$ as a proposal to add to the Markov chain. The proposal is accepted with probability 
\begin{linenomath*}
\begin{align}
P = \min\left(1, \frac{e^{-H(\theta(\tau), p(\tau))}}{e^{-H(\theta(0), p(0))}}\right)\;.
\end{align}
\end{linenomath*}
By energy conservation an analytic solution to the Hamilton equations should always yield probability 1. However, a rejection step is necessary because we solve these equations numerically.
If $\theta(\tau)$ is rejected, then $\theta(0)$ is added to the Markov chain instead. In the limit of an infinite number of samples, the Markov chain converges to samples of the distribution $\hat{L}(\theta)$. 
We seed the Markov chain with a random positive sample from the dataset used to train the neural network.
We set hard walls at the boundary of parameter space by setting the potential energy to infinity.

\subsection{Normalizing Flows}

It is difficult to draw samples from a complicated distribution, like the high-dimensional parameter spaces of the cMSSM and pMSSM. On the other hand, it is easy to draw samples from an equally high-dimensional Gaussian distribution. Normalizing flows is a technique that learns an invertible map $f$ from the simple distribution $p_Z$ to the challenging distribution $p_Y$. One then creates a set of samples from the challenging distribution by mapping easy-to-generate samples:
\begin{linenomath*}
\begin{align}
p_Y(y) = p_Z(f^{-1}(y))\left|\det\left(\frac{\partial f}{\partial y}\right)\right|^{-1}.
\end{align}
\end{linenomath*}
The function $f$ depends on a set of parameters $\Theta$ which are learned by maximizing the log likelihood of a training set, $\mathcal{X}$. The loss function for this training is thus
\begin{linenomath*}
\begin{align*}
\mathcal{L}(\mathcal{X}) &= -\sum_{y \in \mathcal{X}} \left(\log\left(p_Z(f^{-1}(y))\right) - \log\left|\det\left(\frac{\partial f}{\partial y}\right)\right|\right).
\end{align*}
\end{linenomath*}
We construct $f$ to be the composition of $n$ successive maps, $f=f_n\circ\cdots\circ f_1$~\cite{mlapr}. 
Defining $z_{i+1} = f_i(z_i)$ and identifying $y = z_{n+1}$ yields the loss function
\begin{linenomath*}
\begin{align*}
\mathcal{L}(\mathcal{X}) &= -\sum_{y \in \mathcal{X}} \left(\log\left(p_Z(z_1)\right) - \sum_{i=1}^n \log\left|\det\left(\frac{\partial z_{i+1}}{\partial z_{i}}\right)\right|\right).
\end{align*}
\end{linenomath*}
We choose the $f_i$ to be autoregressive transformations. This means that the parameters $\Theta^k_i$ that define the function $f_i$ acting on the $k^\text{th}$ feature $z_i^k$ depends only on the first $(k-1)$ features $z_i^1, \cdots, z_i^{k-1}$. Explicitly,
\begin{linenomath*}
\begin{align*}
z_{i+1}^k = f_i\bigl(z_i^k \,;\; \Theta_i^k(z_{i}^{1:k-1})\bigr).
\end{align*}
\end{linenomath*}
This structure ensures that the Jacobian matrix $\partial z_{i+1}/\partial z_{i}$ is lower triangular so that the determinant is simply the product of diagonal elements. Thus the function $f$ may be efficiently constructed in linear time.

The function $\Theta_i^k\left(z_{i}^{1:k-1}\right)$ can be represented efficiently with a Masked Autoencoder for Distribution Estimation (MADE) \cite{papamakarios2018masked}. MADE networks turn off specific internal weights of the neural network so that the autoregressive property is enforced, allowing one neural network to output all model parameters rather than performing a sequential loop over features.

For our application, we choose $f_i$ to be rational-quadratic neural spline flows with autoregressive layers \cite{durkan2019neural}. These are piece-wise monotonic functions defined as the ratio of two quadratic functions on the interval $[-B, B]$, with $K+1$ knots determining the boundaries between bins. Outside of this interval, the transformation is defined to be the identity. These transformations are parameterized by $3K-1$ parameters for each feature, which are $K$ bin heights, $K$ bin widths, and $K-1$ positive derivative values at the knots, as the derivatives are set to $1$ at $-B$ and $B$ to ensure a continuous derivative over the domain. Permutation layers are included between rational-quadratic transformation layers. We implement the normalizing flow using the Python package \textit{nflows} \cite{durkan2019neural}.

\section{Results}

We analyze the performance of these generative frameworks on the \ac{cMSSM} and \ac{pMSSM} datasets described above. The \ac{cMSSM} is low dimensional and can be scanned relatively well with brute-force search. Thus, we view the \ac{cMSSM} as a test for the generation methods and the \ac{pMSSM} as a more practical application. We present the results for the neural network with \ac{HMC} as well as the normalizing flow side by side. For each method, we generate a dataset of $4\times 10^5$ datapoints. 

We present histograms of generated variables at the \ac{GUT} scale to confirm that the distribution of theory parameters is not biased by our generative framework. We also present histograms of $m_h$ to ensure that our generative models sample within the band of permitted Higgs masses and $\Omega_{\textrm{DM}}h^2$ to provide evidence that the distribution of weak scale quantities match, as these are sensitive to higher order correlations in \ac{GUT} scale parameters.  Finally, we report sampling efficiencies, which are defined as the fraction of the dataset that satisfy a constraint. The hyperparameters used for the supervised neural network, Hamiltonian Monte Carlo, and normalizing flow are given in the Appendix for both datasets.

\subsection{cMSSM}
\label{sec:cMSSM}

\begin{figure*}
\centering
\includegraphics[width=.85\textwidth]{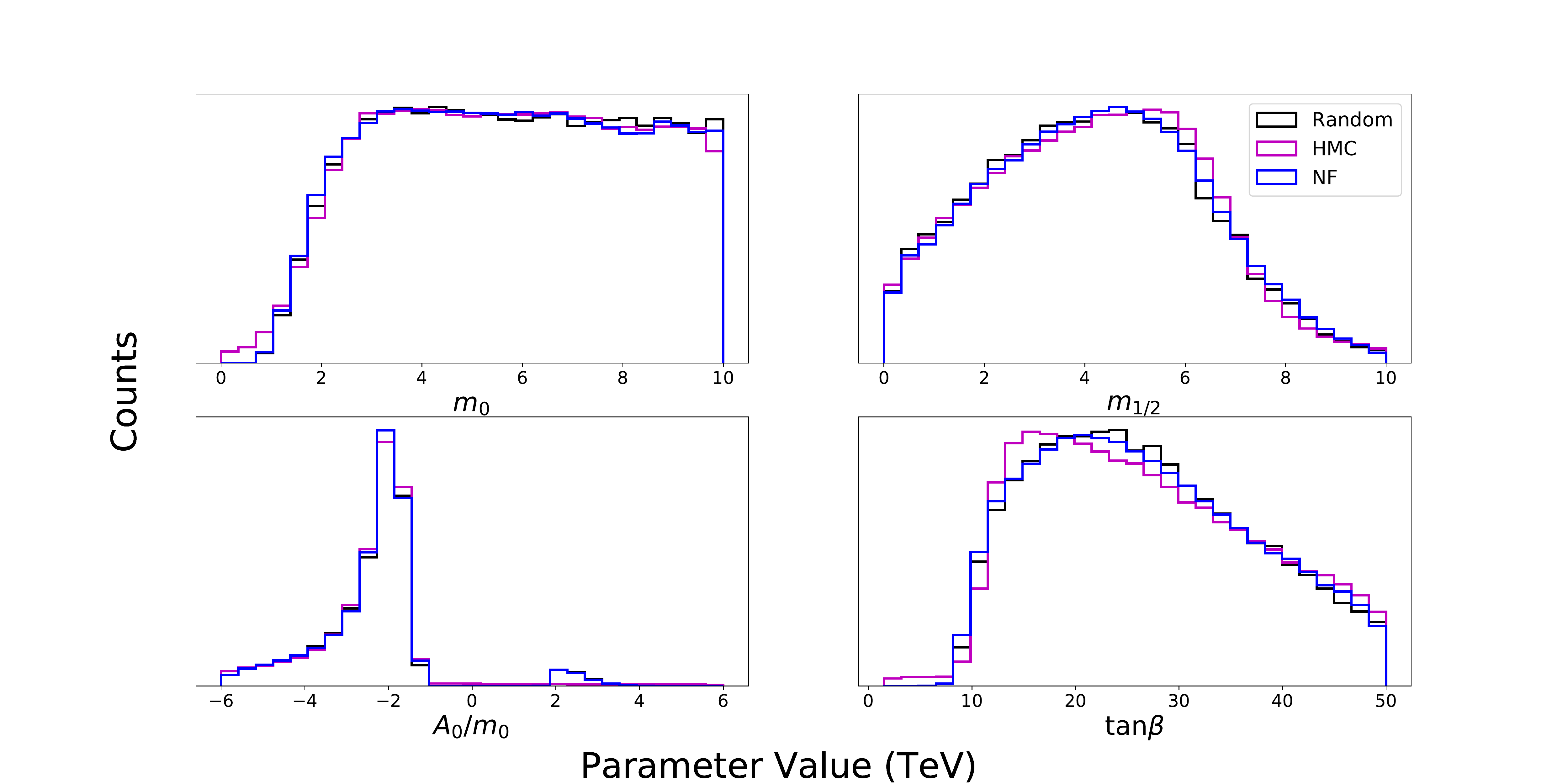}
\caption{Histograms of \ac{cMSSM} parameters that yield the experimental Higgs mass. We observe good agreement between the random sampling, \ac{HMC}, and the flow model. \textsc{Black}: Data obtained through random sampling with a uniform prior and rejecting points that do not have a consistent Higgs mass. \textsc{Magenta}: data sampled with \ac{HMC}. \textsc{Blue}: data sampled from the flow model. No rejection step is applied to generated samples.\label{cmssm_gut}}
\end{figure*}

\begin{figure*}
\centering
\includegraphics[width=.8\textwidth]{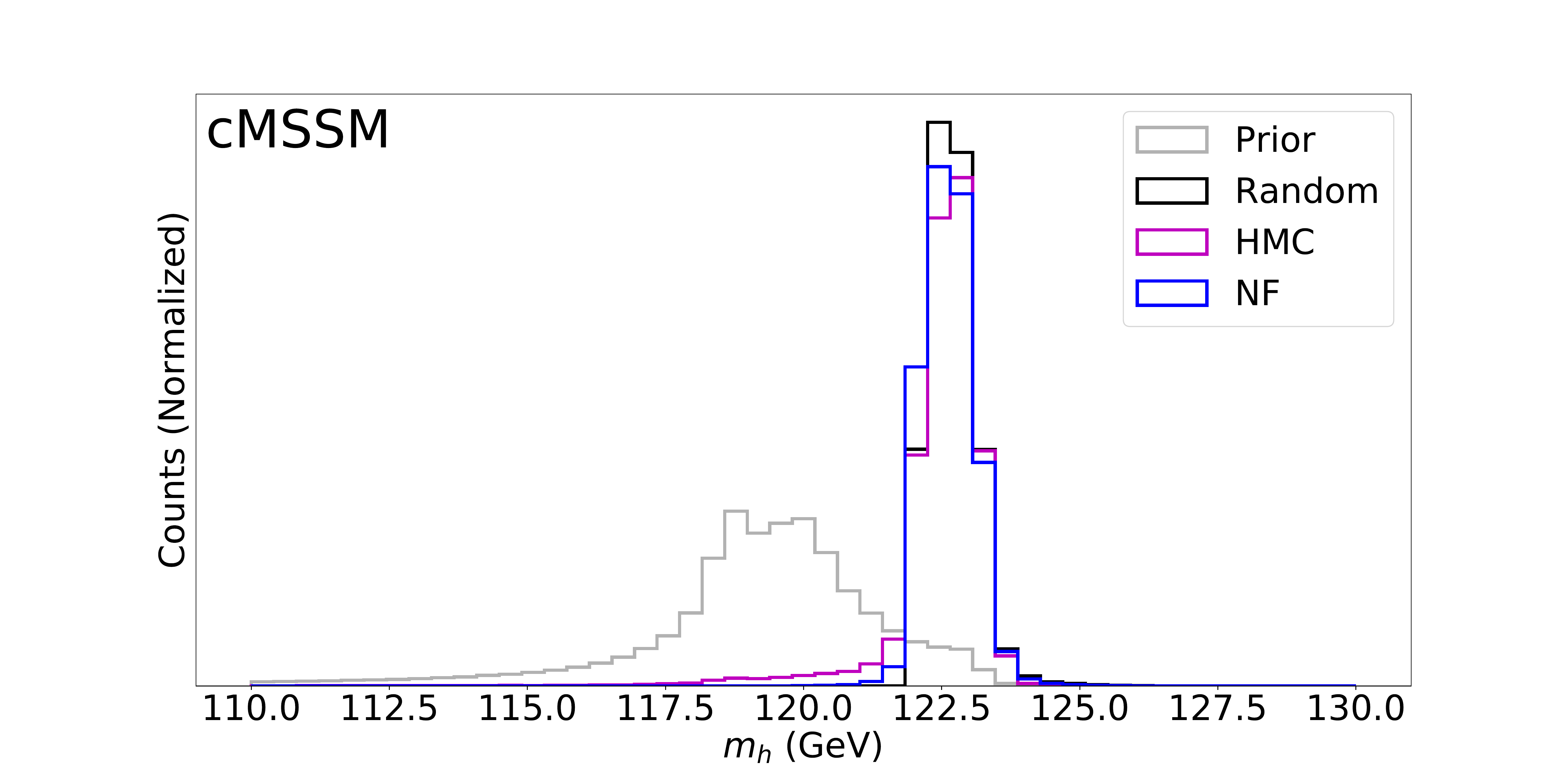}
\caption{Histogram of Higgs masses in the \ac{cMSSM} for different sampling methods. The generative models are seen to mostly sample points consistent with the Higgs mass constraint. \textsc{Gray}: data obtained through random sampling with a uniform prior. \textsc{Black}:  the same randomly sampled data, but points that do not have a consistent Higgs mass are rejected. \textsc{Magenta}: data sampled with HMC. \textsc{Blue}: data sampled with the normalizing flow.\label{cmssm_higgs}}
\end{figure*}

In Figure \ref{cmssm_gut}, we compare histograms of \ac{GUT} scale \ac{cMSSM} parameters. For both generative models, we see very good agreement between the distribution of generated samples and the distribution of randomly sampled points after the Higgs mass constraint is applied. Next, we run the parameters to the weak scale in order to perform the combined search for $\Omega_{\textrm{DM}}h^2$ and $m_h$. In Figure \ref{cmssm_higgs}, we show the distribution of Higgs masses for generated points and randomly sampled points with a rejection step applied. We see that the generative models typically sample within the band of permitted Higgs masses. 

\begin{figure*}
\centering
\includegraphics[width=.8\textwidth]{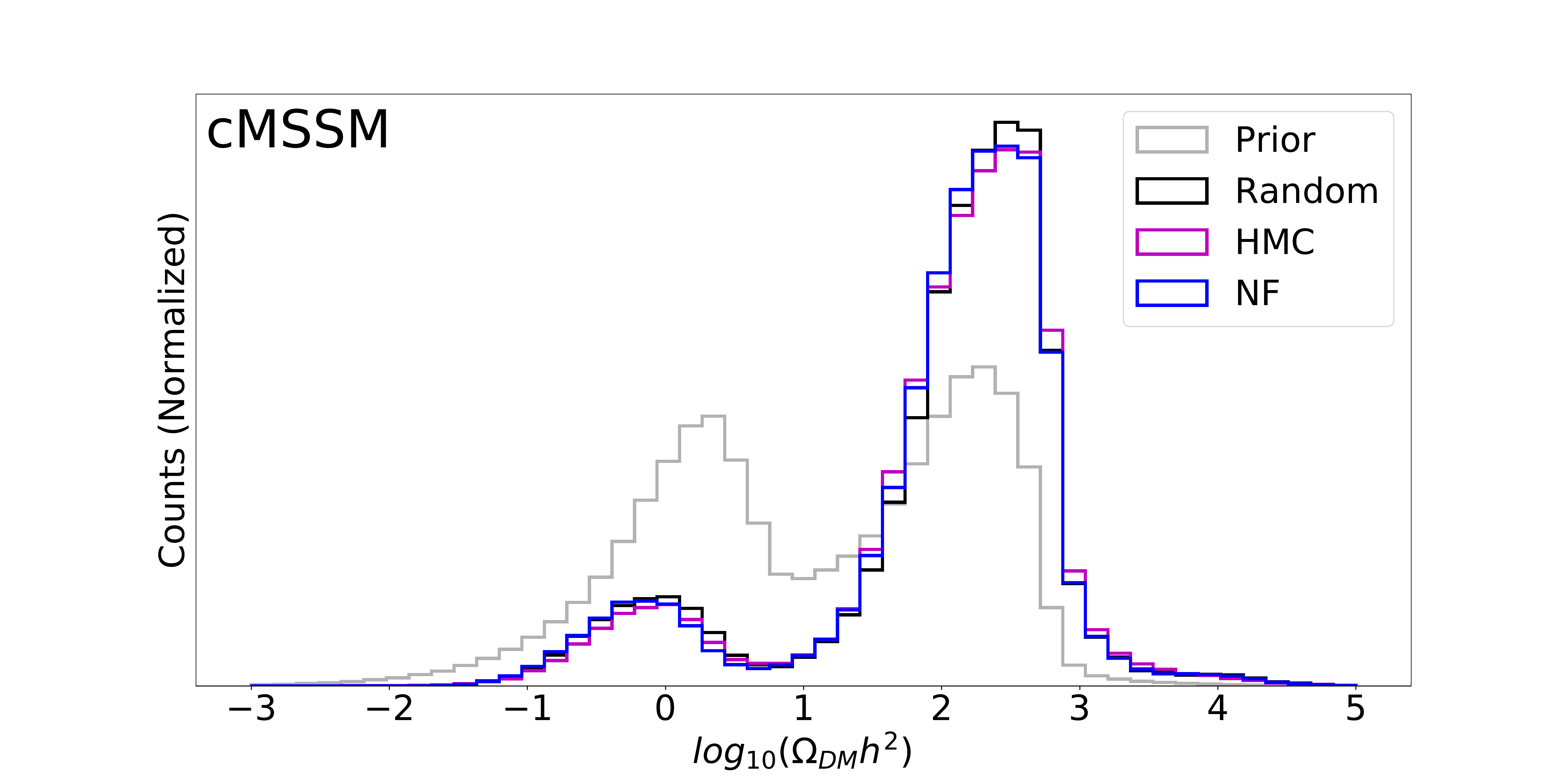}
\caption{Histogram of dark matter thermal relic densities in the \ac{cMSSM} for different sampling methods. We observe that the distributions of the generative models match the distribution of random sampling, providing evidence that the generative models are able to match higher order correlations in \ac{GUT} scale parameters. \textsc{Gray}: data obtained through random sampling with a uniform prior. \textsc{Black}:  the same randomly sampled data, but points that do not have a consistent Higgs mass are rejected. \textsc{Magenta}: data sampled with HMC. \textsc{Blue}: data sampled with the normalizing flow. Generative models have been trained to satisfy the Higgs mass constraint.\label{cmssm_relic_densities}}
\end{figure*}

As an example application in this context, we show histograms of the dark matter relic density for these datasets in Figure \ref{cmssm_relic_densities}. We see that the distribution over dark matter relic densities from the generative models appear to accurately reflect the same distribution in the dataset after the Higgs mass constraint is applied. We emphasize that because the \acp{RGE} are coupled, weak-scale quantities are generally sensitive to higher order correlations of the \ac{GUT} scale parameters, and so matching weak-scale distributions is evidence of matching higher order correlations in the \ac{GUT} scale parameters. This indicates that the $m_h$-constrained subspace has been accurately sampled, allowing for an exploration of additional constraints, such as relic density.

In Table \ref{cmssm_stats}, we compare various statistical properties of random sampling to those of our generative frameworks trained to satisfy the Higgs mass constraint. The first row shows the sampling efficiency with respect to the theoretical constraints mentioned in Section II.A. We see that samples from the generative models are more likely to pass these constraints, as points with a consistent Higgs mass necessarily satisfy the theoretical constraints. The second row shows the sampling efficiency with respect to the Higgs mass constraint. Predictably, the generative models have significantly higher sampling efficiencies than random sampling. We also see that the flow model slightly outperforms the \ac{HMC} sampling method. 

The third row shows the sampling efficiencies with respect to the combined Higgs mass and relic density constraint, where the generative models are still trained to only satisfy the Higgs mass constraint. As mentioned in the introduction, this simulates a scenario where one would like to study the effect of imposing a new constraint in addition to constraints that are already accounted for. Once again, we see that the generative models have much higher sampling efficiencies, resulting from the high probability that the samples pass the Higgs mass constraint. We see an increase in sampling efficiency of approximately an order of magnitude for both generative frameworks.

\begin{table}
\centering
\begin{tabular}{ l l  l  l} \toprule
& \multicolumn{3}{c}{Sampling Method}\\ \cline{2-4}
Constraint & Random & \ac{HMC}$_{m_h}$ & NF$_{m_h}$ \\ \midrule
$\text{Theory}$   & $0.595$ & $0.859$ & $0.879$ \\
$\text{Theory} \cap m_h$  & $0.0389$ & $0.723$ & $0.796$ \\
$\text{Theory} \cap m_h \cap \Omega_{\textrm{DM}}h^2$  & $0.000222$ & $0.00271$ & $0.00456$ \\
\bottomrule
\end{tabular}
\caption{ Comparison of sampling efficiency in the \ac{cMSSM} for several methods and several levels of constraints.  We compare a brute force random scan (random), Hamiltonian MC of a neural network trained to learn the $m_h$ constraint (\ac{HMC}$_{m_h}$), and normalizing flows that incorporate the $m_h$ constraint (NF$_{m_h}$). The constraints applied are theoretical consistency checks (see text), consistency with the experimental Higgs mass and consistency with the Higgs mass and the dark matter relic density $(\Omega_{\textrm{DM}}h^2)$.}\label{cmssm_stats}
\end{table}

\begin{figure*}
\centering
\includegraphics[width=0.85\textwidth]{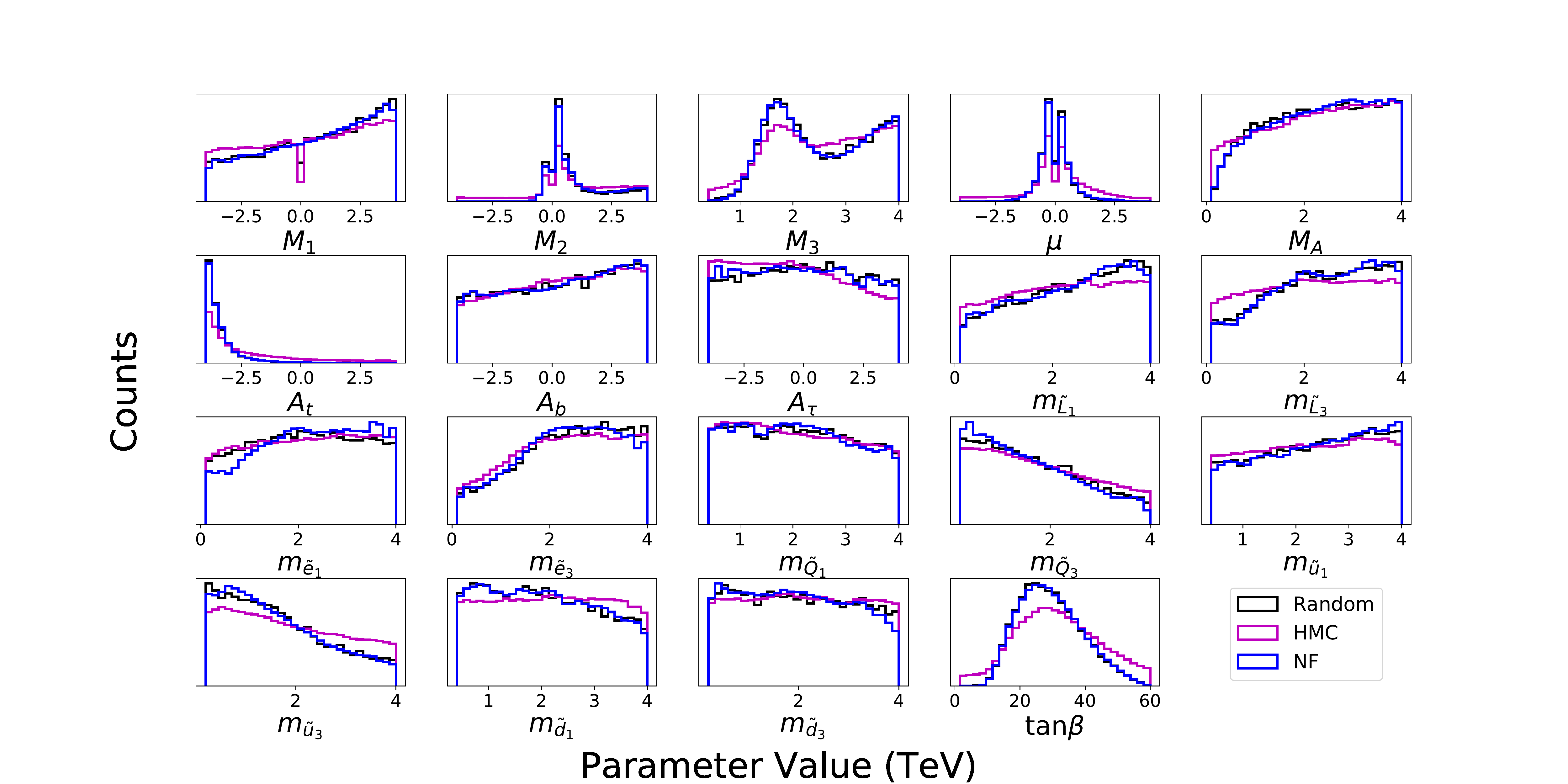}
\caption{Histograms of \ac{pMSSM} parameters that yield the experimental Higgs mass. We observe good agreement between random sampling, \ac{HMC}, and the flow model. \textsc{Black}: Data obtained through random sampling with a uniform prior and rejecting points that do not have a consistent Higgs mass. \textsc{Magenta}: data sampled with \ac{HMC}.
\textsc{Blue}: data sampled from the flow model. No rejection step is applied to generated samples.\label{pmssm_gut}}
\end{figure*}

\subsection{pMSSM}
Differences between the generative models appear in the higher-dimensional \ac{pMSSM}. 
In Figure \ref{pmssm_gut}, we compare histograms of \ac{GUT} scale parameters sampled using brute-force search, \ac{HMC} and the normalizing flow model. 
Despite the increased dimensionality, we find very good agreement in the distributions of all parameters at the \ac{GUT} scale, though HMC has noticeable deviations in some parameters.

\begin{figure*}
\centering
\includegraphics[width=.8\textwidth]{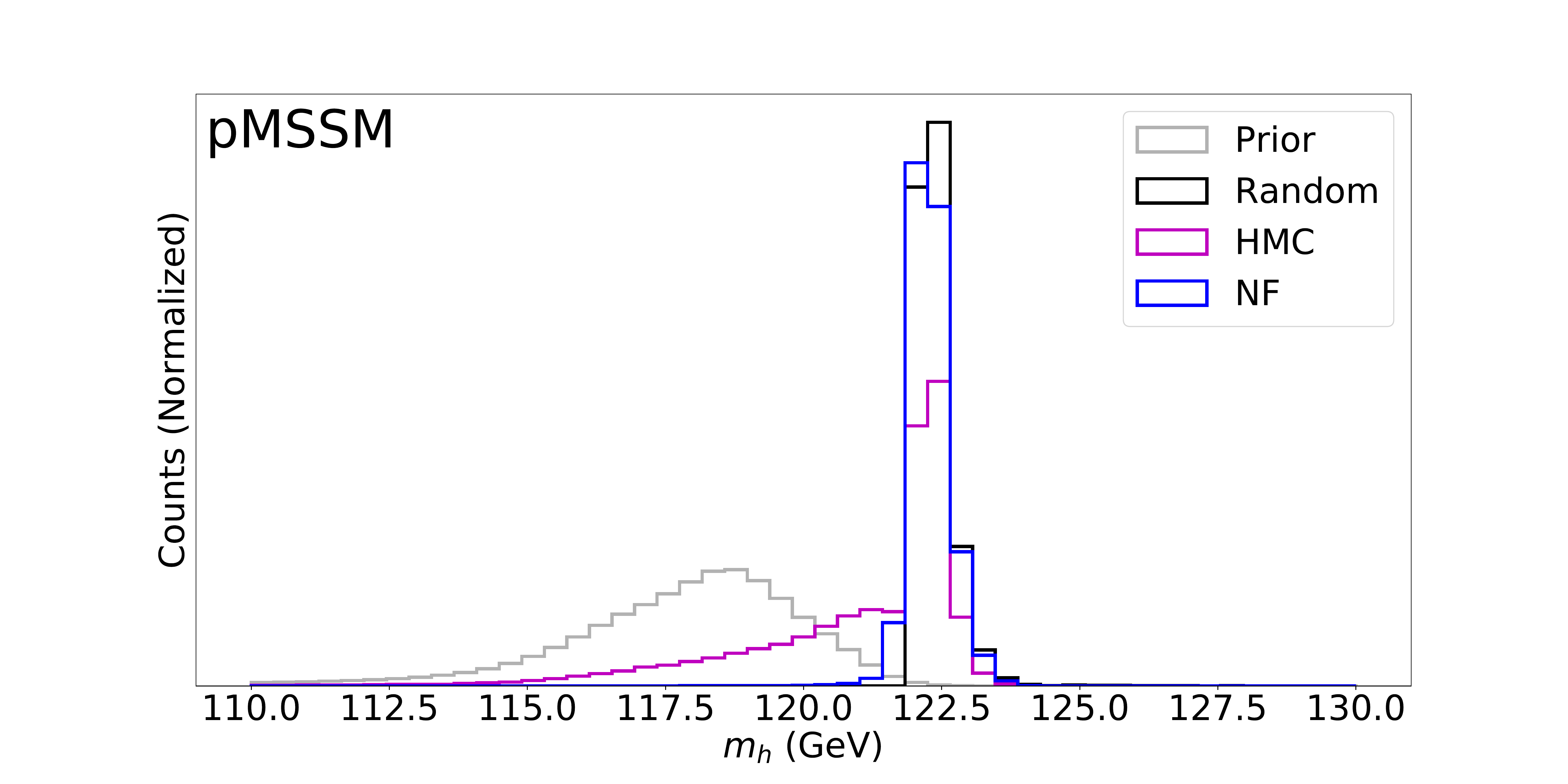}
\caption{Histogram of Higgs masses in the \ac{pMSSM}. The generative models are seen to mostly sample points consistent with the Higgs mass constraint. \textsc{Gray}: data obtained through random sampling with a uniform prior. \textsc{Black}:  the same randomly sampled data, but points that do not have a consistent Higgs mass are rejected. \textsc{Magenta}: data sampled with HMC. \textsc{Blue}: data sampled with the normalizing flow.\label{pmssm_higgs}}
\end{figure*}

\begin{figure*}
\centering
\includegraphics[width=.8\textwidth]{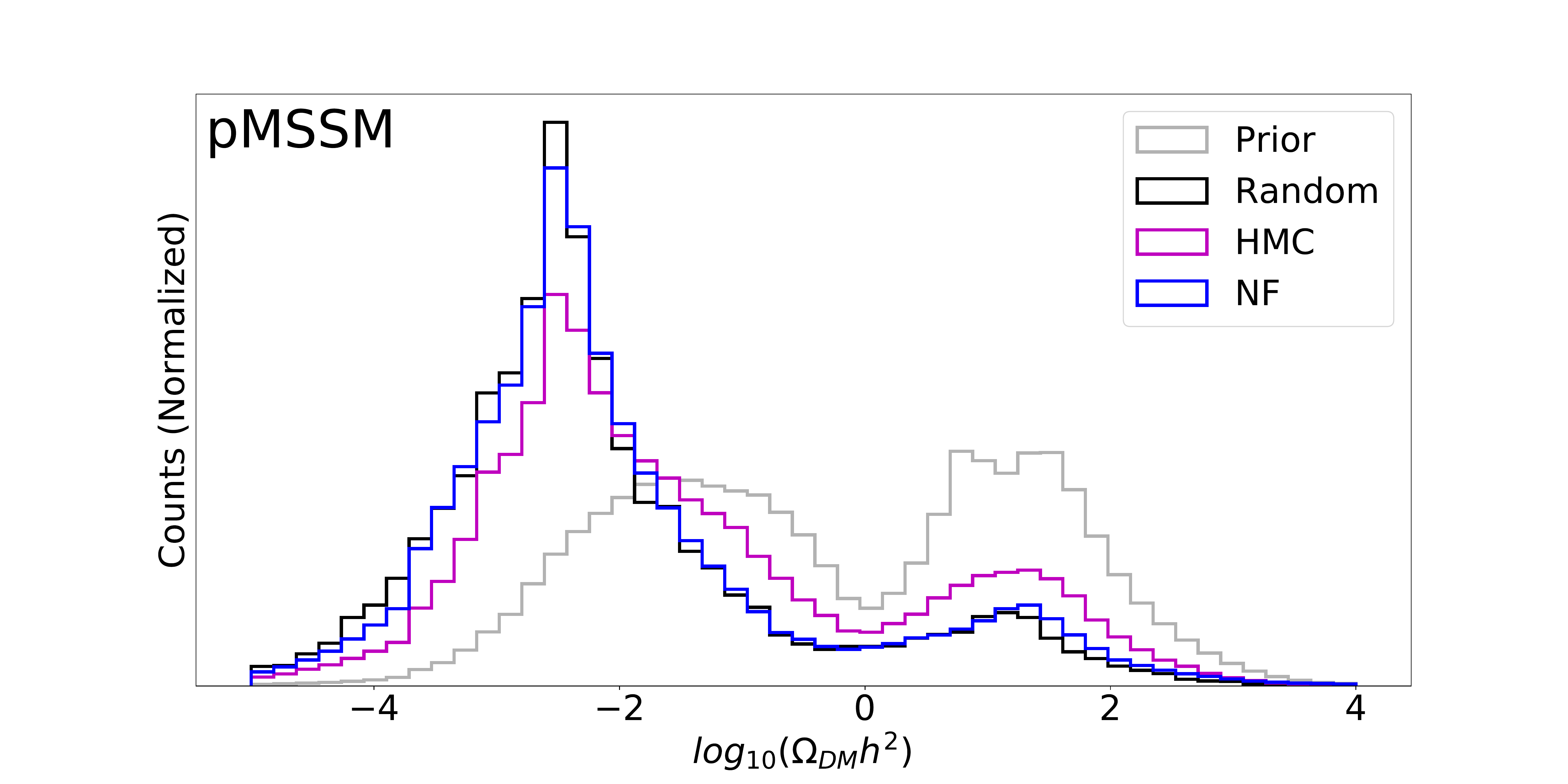}
\caption{Histogram of dark matter thermal relic densities in the \ac{pMSSM}. We observe that the distributions of the generative models match the distribution of random sampling, providing evidence that the generative models are able to match higher order correlations in \ac{GUT} scale parameters. \textsc{Gray}: data obtained through random sampling with a uniform prior. \textsc{Black}:  the same randomly sampled data, but points that do not have a consistent Higgs mass are rejected. \textsc{Magenta}: data sampled with HMC. \textsc{Blue}: data sampled with the normalizing flow. Generative models have been trained to satisfy the Higgs mass constraint.\label{pmssm_relic_densities}}
\end{figure*}

Figures~\ref{pmssm_higgs} and~\ref{pmssm_relic_densities} present histograms of $m_h$ and $\Omega_{\textrm{DM}}h^2$ for the \ac{pMSSM}. 
The generative models tend to sample in the band of allowed Higgs masses, with the normalizing flow model matching the brute-force scan well. The \ac{HMC} samples have a long tail outside of this region towards smaller Higgs masses. We see general agreement with the true distribution of dark matter abundances for both generative frameworks, though the \ac{HMC} samples do not match the brute-force distributions as well as those from the flow model.

Table \ref{pmssm_stats} summarizes the performance of our sampling methods in the \ac{pMSSM}. See Section~\ref{sec:cMSSM} for a detailed description of the quantities presented in the table. Despite the higher dimensionality of the \ac{pMSSM} over the \ac{cMSSM}, we find that generative models increase the sampling efficiency relative to a brute-force search by over two orders of magnitude.
This is much greater than the increase seen in the \ac{cMSSM} and is largely due to the poorer performance of a brute-force search in this space.

\begin{table}
\centering
\begin{tabular}{ l l l l} \toprule
& \multicolumn{3}{c}{Sampling Method}\\ \cline{2-4}
Constraint & Random & \ac{HMC}$_{m_h}$ & NF$_{m_h}$ \\ \midrule
$\text{Theory}$  & $0.553$ & $0.744$ & $0.895$ \\
$\text{Theory} \cap m_h$  & $0.00095$ & $0.319$ & $0.663$ \\
$\text{Theory} \cap m_h \cap \Omega_{\textrm{DM}}h^2$  & $0.000022$ & $0.00574$ & $0.0141$ \\
\bottomrule
\end{tabular}
\caption{Comparison of sampling efficiency in the \ac{pMSSM} for several methods and several levels of constraints.  Methods compared are brute force random scan, Hamiltonian MC of a neural network trained to learn the $m_h$ constraint (\ac{HMC}$_{m_h}$), and normalizing flows that incorporate the $m_h$ constraint (NF$_{m_h}$). Constraints applied are theoretical consistency checks (see text), consistency with the experimental Higgs mass and consistency with the Higgs mass and the dark matter relic density $(\Omega_{\textrm{DM}}h^2)$.\label{pmssm_stats}}
\end{table}

\section{Conclusion}

We implement two generative frameworks that utilize machine learning in order to increase the sampling efficiency of searches in supersymmetric parameter spaces. These sampling methods offer a more efficient way to search the high-dimensional parameter spaces in models of new particle physics. We compare these generative frameworks to the currently used method of a brute-force search, and have seen orders of magnitude of improvement in the sampling efficiency for both parameter spaces considered here. We show that our generative frameworks are able to sample the underlying data distribution without any evidence of bias or mode collapse.

In the \ac{cMSSM}, both methods significantly outperformed random sampling, with the flow model  slightly outperforming \ac{HMC}. In the \ac{pMSSM} the flow model significantly outperforms \ac{HMC}. This is likely due to the larger dimensionality of the \ac{pMSSM}. In addition to performance benefits, the flow model is also quicker to train and sample, making it clearly favorable to \ac{HMC}. However, the \ac{HMC} framework is more complementary to previous works, as it learns the forward problem of determining likelihoods and uses tested Monte Carlo algorithms to sample this likelihood.

Possibilities for future work include incorporating additional constraints into the generative model. In theory, there is no limit to the number of constraints that can be included into either generative model. However, forming an initial dataset for learning may be difficult when the constraints are very strict. A possible remedy is to train generative models with less restrictive constraints which are then used to produce sizable datasets of points that already satisfy many constraints. This new dataset could then be searched to form a training set for a generative model with increasingly restrictive constraints.

Given the ability of the generative machine learning models to efficiently explore high-dimensional parameter spaces, it will be interesting to apply the techniques described here to other problems. For instance, one may identify relations that explain why there is a `little hierarchy' between the electroweak scale and the scale of soft parameters, which go beyond the focus point scenario \cite{Feng:1999zg}. In general, one may be able to identify manifolds of viable points in high-dimensional parameter sets, and explore their geometry.

We have shown promising results in subspaces of the \ac{MSSM} parameter space. These results apply generally to any high-dimensional parameter space with constraints that are computationally expensive to verify. Another direction for future study may be applications to the parameter spaces of even higher-dimensional models of new physics. This includes potentially relaxing constraints built into the pMSSM parameter space, but could also include applications to non-\ac{SUSY} theories. Finally, one could attempt to further tune the neural network structure and hyperparameters in order to achieve higher sample efficiency than was achieved in this work.

\section{Acknowledgements}
The authors would like to thank Tim Cohen, Syris Norelli, Stephan Mandt and Babak Shahbaba. This material is based upon the work supported by the National Science Foundation Graduate Research Fellowship under Grant No. DGE-1321846. DW is supported by the Department of Energy Office of Science. PT is supported by DOE grant DE-SC/0008541. The work of MR is supported by the National Science Foundation under Grant No.\ PHY-1915005.

\section{Appendix}

We present the hyperparameters for our machine learning models in Table~\ref{tab:hyperparams}.

\begin{table}[h]
\centering
\begin{tabular}{ @{\,} >{\raggedright\arraybackslash}p{5mm}p{28mm} p{20mm} p{20mm} @{\,} } \toprule 
	& Parameter & \ac{cMSSM} & \ac{pMSSM}
	\\ 
	\midrule
	\multirow{8}{*}{\rotatebox[origin=c]{90}{Supervised NN} }
		& Learning rate  
		& $0.001$ 
		& $0.0001245$
	\\
    & Hidden layers       & $5$ 	&	$10$ \\
    & Nodes per layer     & $49$ 	&	$154$ \\
    & Dropout             & $0.5$ 	&	$0.0$ \\
    & Activation function & Sigmoid &	Sigmoid \\
    & Optimizer           & ADAM 	&	ADAM \\
    & Batch size          & $128$ 	&	$128$ \\
    & Epochs              & $50$		&	$50$
	\\
	\midrule
	\multirow{6}{*}{\rotatebox[origin=c]{90}{\vphantom{p}HMC}}
		& Step size  \hspace{.2 cm}
		& $0.025$ 
		& $0.008$
	\\
    & Number of steps     & $12$ 		& $12$ \\
    & Mass                & $1.0$ 		& $1.0$ \\
    & Chain length        & $5000$ 		& $5000$ \\
    & Burn-in steps       & $1000$ 		& $1000$ \\
    & Number of chains    & $100$		& $100$
	\\
	\midrule
	\multirow{5}{*}{\rotatebox[origin=c]{90}{\vphantom{p}NF}}
		& Num transforms  
		& $3$ 
		& $3$ 
	\\
    & Batch size          & $1024$ 		& $1024$ \\
    & Epochs              & $300$ 		& $300$ \\
    & $B$                 & $2.0$ 		& $2.0$ \\
    & NN hidden features  & $64$		& $64$ 
	\\
	\bottomrule
\end{tabular}
\caption{\raggedright Hyperparameters used for the machine learning models for to the \ac{cMSSM} and \ac{pMSSM} datasets.} \label{tab:hyperparams}
\end{table}

\newpage
\bibliography{mssm_sampling.bib}

\begin{acronym}
  \acro{MSSM}{Minimal Supersymmetric Standard Model}
  \acro{pMSSM}{phenomenological MSSM}
  \acro{cMSSM}{constrained MSSM}
  \acro{SUSY}{supersymmetry}
  \acro{GUT}{Grand Unified Theory}
  \acro{HMC}{Hamiltonian Monte Carlo}
  \acro{RGE}{renormalization group equation}
\end{acronym}

\end{document}